\documentclass[aps,prb,twocolumn,superscriptaddress,showpacs]{revtex4-1}

\usepackage{amsmath}
\usepackage{amsfonts}
\usepackage{kbordermatrix}
\usepackage{bm}
\usepackage[utf8]{inputenc}
\usepackage[T1]{fontenc}
\usepackage{graphicx}
\usepackage{color}
 \usepackage{ulem}

\newcommand{\conj}[1]{ {#1}^* } %complex number conjugate
\newcommand{\adj}[1]{ {#1}^\dagger } %adjungate
\newcommand{\bra}[1]{ {\left<{#1}\right|} } %bra vector
\newcommand{\cat}[1]{ {\left|{#1}\right>} } %cat vector
\renewcommand{\Re}{ {\operatorname{Re}} }
\renewcommand{\Im}{ {\operatorname{Im}} }

\DeclareMathOperator{\sign}{sign}
\renewcommand{\vec}[1]{ \bm{#1} }
\newcommand{\mat}[1]{ \mathbf{#1} }

\begin{document}

%\title{Spin relaxation in inversion symmetric materials in the strong momentum scattering limit}
\title{An intuitive approach to the unified theory of spin-relaxation}

\author{L\'{e}n\'{a}rd Szolnoki}
\affiliation{Department of Physics, Budapest University of Technology and Economics, POBox 91, H-1521 Budapest, Hungary}
\affiliation{MTA-BME Lend\"{u}let Spintronics Research Group (PROSPIN), Budapest, Hungary}

\author{Bal\'{a}zs D\'{o}ra}
\affiliation{Department of Theoretical Physics, Budapest University of Technology and Economics, PoBox 91, H-1521 Budapest, Hungary}
\affiliation{MTA-BME Lend\"{u}let Spintronics Research Group (PROSPIN), Budapest, Hungary}

\author{Annam\'aria Kiss}
\affiliation{Institute for Solid State Physics and Optics, Wigner Research Centre for Physics, Hungarian Academy of Sciences, POBox 49, H-1525 Budapest, Hungary}
\affiliation{MTA-BME Lend\"{u}let Spintronics Research Group (PROSPIN), Budapest, Hungary}

\author{Jaroslav Fabian}
\affiliation{Institute for Theoretical Physics, University of Regensburg, 93040 Regensburg, Germany}

\author{Ferenc Simon\email{f.simon@eik.bme.hu}}
\affiliation{Department of Physics, Budapest University of Technology and Economics, POBox 91, H-1521 Budapest, Hungary}
\affiliation{MTA-BME Lend\"{u}let Spintronics Research Group (PROSPIN), Budapest, Hungary}

%\date{\today}

\begin{abstract}
Spin-relaxation is conventionally discussed using two different approaches for materials with and without inversion symmetry. The former is known as the Elliott-Yafet (EY) theory and for the latter the D'yakonov-Perel' (DP) theory applies, respectively. We discuss herein a simple and intuitive approach to demonstrate that the two seemingly disparate mechanisms are closely related. A compelling analogy between the respective Hamiltonian is presented and that the usual derivation of spin-relaxation times, in the respective frameworks of the two theories, can be performed. The result also allows to obtain the less canonical spin-relaxation regimes; the generalization of the EY when the material has a large quasiparticle broadening and the DP mechanism in ultrapure semiconductors. The method also allows a practical and intuitive numerical implementation of the spin-relaxation calculation, which is demonstrated for MgB$_2$ that has anomalous spin-relaxation properties.
\end{abstract}
\maketitle

\section{Introduction}
The emerging possibility to use electron spins as information carriers and storage units (a field known as spintronics \cite{WolfSCI,FabianRMP,WuReview}) calls for the development of the fundamental description of spin-diffusion and spin-relaxation phenomena. The conventional theory of spin relaxation in materials is based on the so-called Elliott-Yafet (EY) theory \cite{Elliott,yafet1963g} and the D'yakonov-Perel' (DP) mechanisms \cite{DyakonovPerelSPSS1972,PikusTitkov} depending on whether the material possesses or not a spatial inversion symmetry, respectively. Most elemental metals are examples for a retained inversion symmetry but zincblende semiconductors (e.g. GaAs), even though have a cubic crystal structure, are examples for a broken spatial inversion. The two theories are also based on different approaches: the EY theory studies momentum scattering induced spin-flip events, whereas the DP theory considers the electron spin precession around built-in magnetic fields, which are due to spin-orbit coupling. Common in both theories is that spin-orbit coupling (SOC) is the primary reason of the spin-flip or spin-rotation but it is a small compared to typical orbital interactions.

The spin-relaxation time, $\tau_{\text{s}}$, depends on the momentum relaxation time, $\tau$, however with the opposite dependence: $\tau_{\text{s}}\propto \tau$ in the EY theory, whereas $\tau_{\text{s}}\propto 1/\tau$ in the DP theory. The different phenomenology is used as a benchmark to classify spin-relaxation in a novel material (e.g. graphene \cite{TombrosNat2007,GuntherodtBilayer,KawakamiPRL2010,KawakamiBilayer,YangPRL2011}) into one or the other mechanisms.

Yafet realized in his seminal paper that extension of the conventional EY theory may be required\cite{YafetPL1983}. He predicted properly that band degeneracies lead to anomalous spin-relaxation, such as that observed in aluminium\cite{BMref37}, which was explained by the so-called spin hot-spot modell\cite{FabianPRL1998,FabianPRL1999}. The next hint, that the spin-relaxation theory is yet incomplete and that analogous behaviors of the EY and DP theories exist, came from experiment. Electron spin relaxation studies were performed on materials with spatial inversion symmetry and a short momentum relaxation time, e.g. MgB$_2$ (Ref. \onlinecite{SimonPRL2001}), a material which displays both phenomenology: $\tau_{\text{s}}\propto \tau$ for temperatures below 150 K and $\tau_{\text{s}}\propto 1/\tau$ for $T>400\,\text{K}$ and an intermediate range in between. Another material with a short $\tau$ and retained inversion symmetry is Rb$_3$C$_{60}$ (Ref. \onlinecite{JanossyPRL1993}) which displays $\tau_{\text{s}}\propto 1/\tau$ in the whole temperature range. The experiments in both materials were successfully explained by the so-called generalized EY theory (GEY) in Refs. \onlinecite{SimonPRL2008} and \onlinecite{DoraSimonStronglyCorrPRL2009}. The conventional DP mechanism is also known to require extensions which result in an EY-like phenomenology in inversion symmetry breaking semiconductors with a very long $\tau$, in high magnetic fields \cite{FabianRMP,PikusTitkov,BurkovBalents}, and in the presence of a large spin-orbit coupling \cite{Iordanskii,Averkiev,Oscillatory_DP_theor,Schneider,Oscillatory_DP_exp1,Oscillatory_DP_exp2,Cosacchi}.

The unification of the two theories was first attempted in Ref. \onlinecite{Boross2013}: it was shown that a common quantum kinetic description based approach, and considering different SOC terms for EY-like and DP-like mechanisms, allows to obtain both the EY and the DP theories in a single and compact expression. However, this involved approach appears to be characteristically different from the mathematics of the original EY and DP theories, therefore the intimate link between the two theories remains hidden.

This paper is intended to shed light on this link and we show that formally an analogy between the DP and EY mechanisms can be found. This allows to derive the results of both mechanisms in the framework of the other. We explicitly show this for the EY and GEY mechanisms: the results of spin-relaxation for these cases is derived from a DP-like approach using the quasiparticle picture combined with internal, spin-orbit coupling related magnetic fields. This enables a unified treatment of spin-relaxation and it allows for a readily programmable tool to study spin-relaxation in metals and semiconductors. Our approach also provides an alternative physical description for the EY mechanism: motional narrowing instead of the usual Fermi golden rule description.

\section{The conventional descriptions of spin-relaxation}

\subsection{The D'yakonov-Perel' theory}

  The D'yakonov-Perel' mechanism describes the spin-relaxation in an inversion-breaking material, which is realized in e.g. zincblende (like GaAs) semiconductors. The lack of inversion symmetry gives rise to built-in, SOC related $\bm{k}$-vector dependent effective magnetic fields, $\bm{B}_{k}$, around which the electron spins precess between two momentum scattering events. The corresponding Larmor frequencies of the precession are $\bm{\Omega}_{k}=\gamma\bm{B}_{k}$, where $\gamma \approx 2\pi\cdot 28\,\text{GHz/T}$ is the electron gyromagnetic ratio.

\begin{figure}[htp]
\begin{center}
\includegraphics[width=.7\columnwidth]{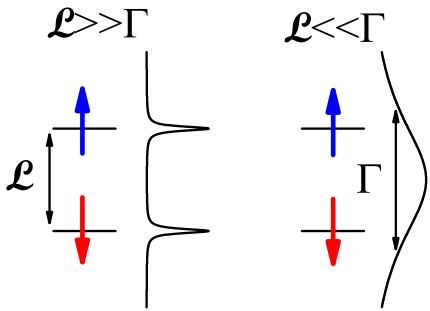}
\caption{Level scheme and the quasiparticle broadening depicted for a material with broken inversion symmetry. Left panel: the situation when $\mathcal{L}\gg \Gamma$, right: level scheme when $\mathcal{L}\ll \Gamma$. The latter is the conventional DP model.}
\label{DP_scheme}
\end{center}
\end{figure}

	The corresponding band structure is shown in Fig. \ref{DP_scheme} for two cases: when momentum scattering is small or large compared to the strength of the spin-orbit interaction. We also introduced in the figure the spin-orbit coupling terms, $\mathcal{L}_{k}$, which correspond to the following Hamiltonian:

	  \begin{equation}
    H_{\text{DP}} =
      \kbordermatrix{
			& \uparrow & \downarrow\\
			\uparrow &0        & \mathcal{L}_{k} \\
      \downarrow&\conj{\mathcal{L}}_{k} & 0 \\},
      \label{eq:H_DP}
  \end{equation}

	\noindent The Hamiltonian is based on spin eigenstates. The connection between the effective magnetic field and the spin-Hamiltonian are: $\hbar\left|\bm{\Omega}_{k}\right|=2\mathcal{L}_{k}$.	In the following, we denote by $\mathcal{L}$ a Fermi surface average, $\left<.\right>_{\text{FS}}$, of the SOC term: $\mathcal{L}=\sqrt{\left<\mathcal{L}_k^2\right>_{\text{FS}}}$. We also introduce the momentum scattering rate, $\Gamma=\hbar/\tau$ and we shall also use the analogous spin-relaxation rate as $\Gamma_{\text{s}}=\hbar/\tau_{\text{s}}$.

The canonical D'yakonov-Perel' theory applies when the momentum relaxation is rapid, $\mathcal{L}\ll \Gamma$, i.e. the electron spins rotate by a small angle $\delta \phi \approx \Omega\tau$ between two consecutive momentum scattering events. The spin-relaxation in this case reads \cite{DyakonovPerelSPSS1972,PikusTitkov}:

\begin{gather}
\Gamma_{\text{s}}=\alpha \frac{\mathcal{L}^2}{\Gamma},
\label{Eq:DPresult}
\end{gather}
i.e. the $\Gamma_{\text{s}}\propto \Gamma^{-1}$ is realized. Herein, $\alpha$ is a band structure dependent constant near unity. The D'yakonov-Perel' mechanism is essentially a motional narrowing description, i.e. the otherwise broad distribution of built-in precession frequencies are averaged out.

		The other limit, i.e. $\mathcal{L}\gg \Gamma$, is realized in ultrapure semiconductors and it was observed experimentally in Refs. \onlinecite{Oscillatory_DP_exp1} and \onlinecite{Oscillatory_DP_exp2} with the corresponding theory worked out in Refs. \onlinecite{Oscillatory_DP_theor,CulcerWinkler,CulcerWinkler2}. We recently studied this problem using a Monte Carlo approach \cite{SzolnokiSciRep} and identified two regimes: if the spread in the Larmor frequencies, $\Delta\bm{\Omega}$, is smaller than $\Gamma/\hbar$, the spin relaxation is oscillatory with a single exponent:

\begin{gather}
		\Gamma_{\text{s}}=\alpha \Gamma.
		\label{Eq:DPwithoscillations}
\end{gather}

		However when $\Delta\Omega\gg\Gamma/\hbar$, the spread in the built-in magnetic fields is larger than $\Gamma$, there is an initial dephasing on the timescale of $1/\Delta\Omega$ that is followed by an exponential-like decay.

 \subsection{The Elliott-Yafet model}

  Our effective Elliott-Yafet model includes 4 bands, two of which remain pairwise degenerate even in the presence of SOC. This is the direct consequence of the Kramers theorem when inversion symmetry is retained, however an external magnetic field would break the time-reversal symmetry and lift the degeneracy. The states in different bands are connected by SOC and the resulting effective Hamiltonian reads:

  \begin{equation}
    H_{\text{EY}} =
      \kbordermatrix{
        & 1\uparrow & 1\downarrow & 2\uparrow & 2\downarrow \\
        1\uparrow   & 0        & 0 & 0        & L_{k} \\
        1\downarrow & 0        & 0 & \conj{L}_{k} & 0 \\
        2\uparrow   & 0        & L_{k} & \Delta_{k}   & 0 \\
        2\downarrow & \conj{L}_{k} & 0 & 0        & \Delta_{k}
      },
      \label{eq:H_EY}
  \end{equation}
  \noindent where $1$ and $2$ are orbital states, $L_{k}$ is the spin-orbit matrix element and $\Delta_{k}$ is the band gap. In principle, there could be matrix elements between $1\uparrow$ and $2\uparrow$ but these would not contribute to spin-relaxation \cite{Boross2013} in leading order and are thus omitted. Similar to the introduction of $\mathcal{L}$, we define the following average values: $L=\sqrt{\left<L_k^2\right>_{\text{FS}}}$ and $\Delta=\sqrt{\left<\Delta_k^2\right>_{\text{FS}}}$.

\begin{figure}[htp]
\begin{center}
\includegraphics[width=.7\columnwidth]{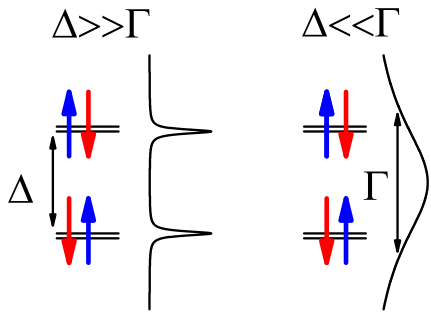}
\caption{Level scheme and the momentum scattering rate depicted for a material with retained inversional symmetry. Note that the spin states are degenerate. Left panel: the situation when $\Delta \gg \Gamma$, right: level scheme when $\Delta \ll \Gamma$. The earlier is the conventional EY model, while the latter is the so-called generalized EY model, worked out in Refs. \onlinecite{SimonPRL2008} and \onlinecite{DoraSimonStronglyCorrPRL2009}.}
\label{EY_scheme}
\end{center}
\end{figure}

In Fig. \ref{EY_scheme}., we show the level scheme for the case when inversion symmetry is retained. The conventional EY model assumes the hierarchy $L\ll \Gamma \ll \Delta$ therefore the states are weak admixtures of the original pure spin-up/down states:

\begin{equation}
\begin{aligned}
|\widetilde{1\uparrow}\rangle_{k}= |1\uparrow \rangle_{k}+\frac{L_{k}}{\Delta_{k}}|2\downarrow \rangle_{k}\\
|\widetilde{1\downarrow}\rangle_{k}= |1\downarrow \rangle_{k}-\frac{\conj{L}_{k}}{\Delta_{k}}|2\uparrow \rangle_{k}.
\end{aligned}
\label{Eq:EYmixed_states}
\end{equation}
and correspondingly for the other two states.

Elliott considered the probability of a spin-flip during momentum scattering between the $|\widetilde{1\uparrow}\rangle_{k}$ and $|\widetilde{1\downarrow}\rangle_{k}$ eigenstates and showed using first order time dependent perturbation theory that the transition probability of the spin-flip, $W_{k\widetilde{\uparrow}\rightarrow k'\widetilde{\downarrow}}$, and momentum scattering, $W_{k\rightarrow k'}$, processes are related by:

\begin{gather}
W_{k\widetilde{\uparrow}\rightarrow k'\widetilde{\downarrow}}=\alpha \frac{L_{k}^2}{\Delta_{k}^2}W_{k\rightarrow k'},
\end{gather}
where $\alpha$ is again a band structure dependent constant near unity. This yields the well known EY expression:

\begin{gather}
\Gamma_{\text{s}}=\left<\frac{L_k^2}{\Delta_k^2}\right>_{\text{FS}}\cdot\Gamma\approx \alpha \frac{L^2}{\Delta^2}\Gamma
\label{Eq:EYresult}
\end{gather}

The so-called generalized EY scenario is depicted in the right panel of Fig. \ref{EY_scheme} with the corresponding hierarchy of the parameters: $L\ll \Delta \ll\Gamma$. This situation is realized in materials where there is a four-fold band degeneracy (such as e.g. in graphene at the Dirac point) or a near band-degeneracy whose splitting is smaller than the momentum scattering rate. The latter situation is encountered in MgB$_2$ which is a relatively high temperature superconductor ($T_{\text{c}}=40\,\text{K}$) thus it has a strong electron-phonon coupling (i.e. a large $\Gamma$) and the conduction band has a near band-degeneracy due to the two boron atoms in the unit cell. The other known examples are alkali atom doped fullerides, K$_3$C$_{60}$ and Rb$_3$C$_{60}$, which also show a large electron-phonon coupling and a conduction band which is derived from the three-fold degenerate C$_{60}$ molecular orbitals and the degeneracy lifting is moderate compared to $\Gamma$.

Clearly, for such a case the lowest order time-dependent perturbation theory is not applicable. A many-body based approach, which is essentially infinite order in momentum scattering (but remains lowest, i.e. second order in the SOC), was performed in Refs. \onlinecite{SimonPRL2008,DoraSimonStronglyCorrPRL2009}. It gave the remarkably simple result:

\begin{gather}
\Gamma_{\text{s}}=\alpha\frac{L^2}{\Delta^2+\Gamma^2}\Gamma.
\label{Eq:GEYresult}
\end{gather}

Clearly, the GEY regime gives: $\Gamma_{\text{s}}\propto \frac{L^2}{\Gamma}$, which looks surprisingly similar to the DP result of Eq. \eqref{Eq:DPresult}. This immediately leads to questions as to whether the agreement is only accidental or there is something behind the agreement, i.e. that a large $\Gamma$ inevitably corresponds to a breaking of the lattice symmetry (both phonons and impurities lead to this).

Another point awaiting clarification is whether a simpler, less involved derivation of the GEY regime is possible without the many-body considerations.

\section{Analogy of the EY and DP Hamiltonians}

We first demonstrate how the EY Hamiltonian can be related to the DP problem. We consider a transformation of the spin states in the Elliott-Yafet Hamiltonian in Eq.~\eqref{eq:H_EY} as follows:

  \begin{equation}
    \begin{aligned}
      \cat{A\uparrow}   &= \cat{1\uparrow} \\
      \cat{A\downarrow} &= \cat{2\downarrow} \\
      \cat{B\uparrow}   &= \cat{2\uparrow} \\
      \cat{B\downarrow} &= \cat{1\downarrow}.
    \end{aligned}
    \label{eq:newstates}
  \end{equation}
  The Hamiltonian for these states reads:

  \begin{equation}
    H'_{\text{EY}} =
      \kbordermatrix{
        & A\uparrow & A\downarrow & B\uparrow & B\downarrow \\
        A\uparrow   & 0        & L_{k}      & 0        & 0 \\
        A\downarrow & \conj{L_{k}} & \Delta_{k} & 0        & 0 \\
        B\uparrow   & 0        & 0      & \Delta_{k}   & L_{k} \\
        B\downarrow & 0        & 0      & \conj{L_{k}} & 0
      }.
    \label{eq:H_EY_renamed}
  \end{equation}

This procedure is equivalent to a unitary transformation of the EY Hamiltionian with the following matrix:

  \begin{equation}
    \begin{aligned}
      H'_{\text{EY}} &= \adj{P} H_{\text{EY}} P, \\
      P &= 
        \begin{bmatrix}
          1 & 0 & 0 & 0 \\
          0 & 0 & 0 & 1 \\
          0 & 0 & 1 & 0 \\
          0 & 1 & 0 & 0 
        \end{bmatrix}.
    \end{aligned}
    \label{eq:Permutation_matrix}
  \end{equation}

In the original basis we had pure spin states, however in the new basis the usual Pauli matrices need to be transformed, too. Therefore we rather refer to the new basis states as pseudo-spin states.

This Hamiltonian decouples to the direct sum of two single pseudospin Hamiltonians. This is the consequence of the simple structure of the original Hamiltonian in Eq. \eqref{eq:H_EY} as it did not contain terms which would join e.g. $A\uparrow$ with $B\uparrow$. It is important to note that momentum scattering does not induce transitions between the $A$ and $B$ states either in this model Hamiltonian as momentum scattering alone is unable to flip an electron spin, except for a magnetic scatterer which is not involved in the EY theory.
%The spin relaxation in the theory is solely due to the mixing of pure spin states in the wavefunction of the material.
This means that if an electron is born to be of the $A$ type, it remains so throughout in the spin-relaxation problem. We discuss it in the Supplementary Information that the procedure can be performed in the presence of terms like between the $1\uparrow$ and $2\uparrow$ states as well but the result lacks compactness.

The 2x2 blocks in Eq. \eqref{eq:H_EY_renamed} show a remarkable similarity to the DP Hamiltonian in Eq. \eqref{eq:H_DP} except for the presence of the $\Delta$ term. We note that the $L_k$ terms are the SOC matrix elements for a material with retained inversion symmetry. As we show below, this allows to "formally" treat the EY spin-relaxation mechanism with the usual DP-like approach, i.e. by the evolution of the pseudospin ensemble upon internal built-in magnetic fields. In the end of the pseudo-spin evolution, the result can be obtained for the real spin by transforming it back. 

We then discuss how the DP Hamiltonian can be traced back to the EY problem. We consider the DP Hamiltonian in Eq. \eqref{eq:H_DP}, introduce an external magnetic field, with Zeeman energy $\Delta_{\text{Z}}$, and amend the system with a pair of virtual electron spin states, which have the same SOC but interact with the external magnetic field with opposite sign as compared to the physical electron states. This gives:

%\kbordermatrix{\\&0        & \mathcal{L}_{k} \\
 %       &\conj{\mathcal{L}}_{k} & 0 \\},

 \begin{equation}
    H'_{\text{DP}} =
      \kbordermatrix{\\
        &0        & \mathcal{L}_{k}      & 0        & 0 \\
        &\conj{\mathcal{L}_{k}} & \Delta_{\text{Z},k} & 0        & 0 \\
        &0        & 0      & \Delta_{\text{Z},k}   & \mathcal{L}_{k} \\
        &0        & 0      & \conj{\mathcal{L}_{k}} & 0
      }.
    \label{eq:H_DP_renamed}
  \end{equation}

We also shifted the zero of the energy as the Zeeman term would normally introduce a $\pm \Delta_{\text{Z}}/2$ shift of the levels. We retained the $k$ dependence of the Zeeman term as the $g$-factor can be $k$-dependent. Clearly, a similar transformation of the states, indentical to that presented above, yields the EY Hamiltonian for the DP problem. We emphasize that the $\mathcal{L}_{k}$ terms originate from an SOC due to inversion symmetry breaking, still it can be rewritten as if it was valid for a system with retained inversion. 

%This amendment of the states with a pair of virtual electron spin states corresponds to restoring the inversion symmetry of the material in real space. E.g. for GaAs it means adding a spatially inverted AsGa counterpart to the original lattice.

Given that the additional virtual electron states are not connected with the true physical electrons, we believe that the usual treatment of spin-relaxation in inversionally symmetric materials, i.e. first order perturbation theory for the conventional EY and many-body approach for the GEY, can be performed for these states, too.

\section{Deriving the EY and GEY results using a DP consideration}

The Hamiltonian in Eq. \eqref{eq:H_EY_renamed} describes independent time evolution for the $A$ and $B$ electrons and their pseudospins, which can be described by Larmor precession with the following angular frequency vectors:

	 \begin{equation}
      \bm\Omega_{A,k} = \frac{2}{\hbar}
        \begin{bmatrix}
          -\Re[L_k] \\ \Im[L_k] \\ \Delta_k/2
        \end{bmatrix},
      \quad
      \bm\Omega_{B,k} = \frac{2}{\hbar}
        \begin{bmatrix}
          -\Re[L_k] \\ \Im[L_k] \\ -\Delta_k/2
        \end{bmatrix}
		\label{eq:Larmor}
  \end{equation}

The presence of an effective model which contains a two pseudospin-level state only with an internal, SOC derived magnetic field is a surprisingly analogous situation to the DP description of spin-relaxation in semiconductors. We emphasize that the original system has inversion symmetry but at least formally it can be rewritten to look like the DP Hamiltonian in Eq. \eqref{eq:H_DP}. It is therefore tempting to employ the DP framework to derive the EY and GEY results. This means that one follows the time evolution of a pseudospin ensemble upon Larmor precessions according to Eq. \eqref{eq:Larmor} between two momentum scattering events. We note that the basis change from $\cat{1\uparrow}...$ to $\cat{A\uparrow}...$ affects the $x,y$ rotations of the original spin, however the transformation leaves the $s_z$ Pauli matrix invariant. In the models of spin-relaxation, the $z$ or a quantization axis is chosen along which a non-equilibrium spin population is induced either by magnetic resonance or by spin injection. This means that we are following the time decay of the $z$ component, which is identical for both the usual spin and the pseudospin.

The spin-relaxation under the action of built-in SOC related fields and an additional magnetic field along the $z$ axis was worked out analytically in Ref. \onlinecite{FabianActaPhysSlovaca} (Eq. IV. 36):

\begin{gather}
\Gamma_{\text{s}}=\left( \overline{\Omega_x^2}+\overline{\Omega_y^2}\right)\frac{\tau_c}{\Omega_0^2\tau_c^2+1}. 
\label{Jaro_formula}
\end{gather}
Herein, $\Omega_x$ and $\Omega_y$ denote the Larmor (angular) frequencies due to SOC fields along the $x$ and $y$ axes, respectively, which in our notation are identified with $\hbar\Omega_x=-2\Re[L_k]$ and $\hbar\Omega_y=2\Im[L_k]$. $\tau_c$ is the correlation time of the electron scattering, which is identified with the usual momentum scattering in our case and $\Omega_0$ is the Larmor angular frequency due to the magnetic field along the $z$ axis. In the EY model, $\Delta_k$ plays the role of the Zeeman splitting, thus the assignment is $\hbar \Omega_0=\Delta$. Clearly, Eq. \eqref{Jaro_formula}, together with the angular frequencies given in Eq. \eqref{eq:Larmor} can be rewritten as:

\begin{gather}
\Gamma_{\text{s}}=\frac{4L^2}{\Delta^2+\Gamma^2}\Gamma
\label{Jaro_formula_rewritten}
\end{gather}

This formula is identical to Eq. \eqref{Eq:GEYresult}, which was originally derived for the generalized EY mechanism using a quantum kinetic approach. 

Let us briefly reflect on the above derivation. We essentially start with pure spin states (Pauli eigenstates) and consider both disorder and spin-orbit coupling as perturbations. The spin-orbit coupling leads to a coherent spin dynamics between two non-degenerate levels,
separated by $\Delta_k$. If the scattering is faster than the spin precession, we are in 
the motional narrowing regime and the EY mechanism and the electron state does not 
become fully "dressed" by SOC. In contrast, if the coherent dynamics is long enough, the electron
states fully explore the Hilbert states of the available bands and the spin relaxation approaches the EY formula. However, the above does not hold in the spin hot-spot regime \cite{FabianPRL1998,FabianPRL1999} where $\Delta_k \alt L_k$. In this regime the spin-flip probability can be as large as the spin-conserving one and perturbation theory within the Markov approximation, which leads to the formula Eq. \ref{Jaro_formula_rewritten} is not valid.

\section{Spin-relaxation rate calculation for a realistic case}

The above presented unified treatment of the EY-GEY and DP spin-relaxation regimes not only represents a simple and intuitive description but it allows for a practical and easily implemented calculation of spin-relaxation times (or rates). The original EY theory and in particular the GEY regime is difficult to implement in practice to yield spin-relaxation rate values. However, the treatment of the DP scenario is more intuitive: it considers the evolution of a spin ensemble under the action of built-in SOC fields. We demonstrate this for MgB$_2$: in this material, a crossover from the conventional EY to the GEY regime was observed experimentally using electron spin resonance measurements \cite{SimonPRL2001,SimonPRL2008}. At low $T$, the conventional EY regime dominates as the ESR linewidth and resistivity are proportional, thus $\Gamma_{\text{s}}\propto \Gamma$. However, above about 200 K, it crosses over to the GEY regime as the quasiparticle broadening, $\Gamma$, becomes comparable to the band-band separation of boron $\sigma$ bands in the vicinity of the Fermi surface. The large $\Gamma$ is the result of a large electron-phonon coupling, which is also the cause of the superconducting transition at 40 K. The presence of nearly degenerate boron $\sigma$ bands on the Fermi surface is due to the presence of two boron basis atoms.

\begin{figure}[htp]
\begin{center}
\includegraphics[width=1\columnwidth]{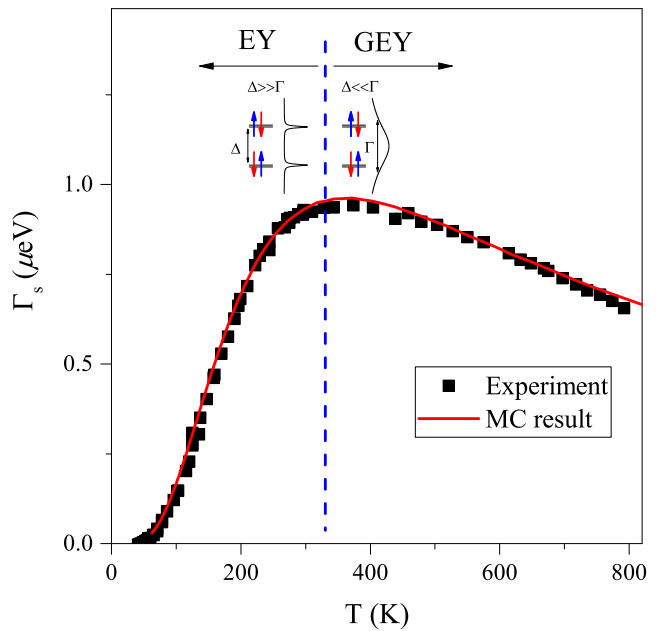}
\caption{Experimental (symbols) and Monte Carlo based (solid curve) spin-relaxation rate, $\Gamma_{\text{s}}$ in MgB$_2$. Note that the experimental data contains the contribution from the boron $\sigma$ bands only; as described in Ref. \onlinecite{SimonPRL2008}, the contribution from the $\pi$ bands remains explicable by the conventional EY theory and it is thus subtracted. The level scheme and the identification of the EY and GEY regimes is also depicted. Note that the two regimes cross over smoothly as a function of temperature and the vertical dashed line is intended as a guide only.}
\label{fig:Fig4_MgB2_calc}
\end{center}
\end{figure}

We show the experimental and calculated spin-relaxation rates, $\Gamma_{\text{s}}$, in Fig. \ref{fig:Fig4_MgB2_calc}. The experimental data are from Ref. \onlinecite{SimonPRL2008} and is obtained after subtracting the spin-relaxation rate contribution from the boron $\pi$ bands. It is described in Ref. \onlinecite{SimonPRL2008} that the latter obeys the conventional EY mechanism as the boron $\pi$ bands are separated from each other with a large gap (2 eV). In contrast, the separation of the boron $\sigma$ bands is as low as $\sim 0.2\,\text{eV}$.

To calculate the spin-relaxation rate in the framework of the above description, we consider the EY Hamiltonian of Eq. \eqref{eq:H_EY} with an isotropic SOC (i.e. the off-diagonal 2x2 matrix have diagonal terms, too). It is transformed to a DP-like Hamiltonian according to the procedure described above (the diagonal terms of the off-diagonal 2x2 matrix are eliminated according to Ref. \onlinecite{Supmat}). The rest of the calculation proceeds according to the Monte Carlo simulation procedure of a DP problem, which is discussed in depth in Ref. \onlinecite{SzolnokiSciRep} and is briefly summarized herein. 

The Monte Carlo method considers an electron ensemble with spin polarized along the $z$ axis at $t=0$. Each electron has a different (random) $\bm{k}$ value and they experience a different built-in Larmor precession. The latter vectors are obtained according to the above described transformation procedure. Momentum scattering is approximated by giving each electron a new random $\bm{k}$ value after a time $\tau'$ elapses, where $\tau'$ is a statistical variable with a Poisson distribution and a mean value of the momentum relaxation time, $\tau$. The decay of the electron spin ensemble magnetization along the $z$ axis is monitored as a function of time and is fitted with a single exponential. The exponent yields directly the spin-relaxation time, $\tau_{\text{s}}$. The variation of this parameter, or its rate $\Gamma_{\text{s}}=\hbar/\tau_{\text{s}}$ is followed as a function of $\Gamma$. Ref. \onlinecite{Supmat} shows $\Gamma_{\text{s}}$ in units of $L^2/\Delta$ as a function of $\Gamma$ (in units of $\Delta$) for a low value of the spin-orbit coupling ($L=0.1\Delta$). In the low $L$ limit, this $\Gamma_{\text{s}}(\Gamma)$ function is universal and can be well fitted by $\Gamma_{\text{s}}=\frac{2}{3}\frac{\Gamma}{1+\Gamma^2}$. This is exactly the generalized EY result of Eq. \ref{Eq:GEYresult} which was first deduced in Ref. \onlinecite{SimonPRL2008} and is obtained herein numerically.

The final result, which is shown in Fig. \ref{fig:Fig4_MgB2_calc}, is obtained by presenting the data as a function of temperature using the $T(\Gamma)$ relation from the Bloch-Gr\"{u}nneisen function with a Debye temperature of 535 K and also by scaling the data with the already known $\Delta=194\,\text{meV}$ (Ref. \onlinecite{SimonPRL2008}) and $L=1.7\,\text{meV}$. 

We observe an excellent agreement between the experiment and the spin-relaxation rate which is obtained from the Monte Carlo simulation. This also means that the spin-relaxation rate in MgB$_2$ can be appropriately described by a single band-band separation value and also by an isotropic SOC model. We emphasize that with this demonstration, we traced back a spin-relaxation problem in an inversionally symmetric strongly correlated metal to the methodology which was developed for an inversionally symmetry breaking material. 

\section{Summary}
In summary, we presented the connection between the Elliott-Yafet and D'yakonov-Perel' theories of spin-relaxation. We showed that a compelling analogy can be realized between the two mechanisms concerning the level structure and the respective model Hamiltonians. This allows an intuitive, albeit not rigorous derivation of both mechanisms based on the framework of the other. We showed recently that the DP-like treatment is readily programmable using a Monte Carlo method \cite{SzolnokiSciRep}. This means that the otherwise involved quantum mechanical approach of the EY theory, and in particular its extension to the case of strong electron phonon coupling (the so-called generalized EY theory, GEY), can be also treated at a Monte Carlo level. This is explicitly demonstrated for MgB$_2$, where the EY to GEY crossover is observed and is reproduced numerically.

\section*{Acknowledgements}
Work supported by the Hungarian National Research, Development and Innovation Office (NKFIH) Grant Nr. K119442, by DFG SFB 1277. A.K. acknowledges the Bolyai Program of the Hungarian Academy of Sciences.

%\bibliographystyle{apsrev}
%\bibliographystyle{naturemag}
%\bibliography{Tubes2011June_new}

%\clearpage

%\onecolumngrid
\appendix

\section{Block diagonalization of the general four-band Elliott-Yafet Hamiltonian}

The most general form of a four-band Hamiltonian that is symmetric to inversion:

  \begin{equation}
    H_{\text{EY}} =
      \kbordermatrix{
                      & 1\uparrow      & 1\downarrow & 2\uparrow      & 2\downarrow \\
          1\uparrow   & 0              & 0           & L_{2,k}        & L_{1,k} \\
          1\downarrow & 0              & 0           & \conj{L}_{1,k} & -L_{2,k} \\
          2\uparrow   & L_{2,k}        & L_{1,k}     & \Delta_{k}     & 0 \\
          2\downarrow & \conj{L}_{1,k} & -L_{2,k}    & 0              & \Delta_{k}
      },
      \label{eq:H_EY_matrix}
  \end{equation}

\noindent where $L_{2,k}$ is real.
After transforming the states similarly to the main article, we get a Hamiltonian that is not complately block-diagonalized:

  \begin{equation}
    H'_{\text{EY}} =
      \kbordermatrix{
                    & A\uparrow      & A\downarrow & B\uparrow      & B\downarrow \\
        A\uparrow   & 0              & L_{1,k}     & L_{2,k}        & 0 \\
        A\downarrow & \conj{L}_{1,k} & \Delta_{k}  & 0              & -L_{2,k} \\
        B\uparrow   & L_{2,k}        & 0           & \Delta_{k}     & L_{1,k} \\
        B\downarrow & 0              & -L_{2,k}    & \conj{L}_{1,k} & 0
      }.
    \label{eq:H_EY_renamed_gen}
  \end{equation}

We can eliminate the $\pm L_{2,k}$ matrix elements by a two dimensional rotation between the kinetic states with a unitary operation which is very similar to the one presented in the main text.

\begin{equation}
  \begin{aligned}
    \cat{A'} &= a \cat{A} - b \cat{B}, \\
    \cat{B'} &= b \cat{A} + a \cat{B}, \\
    a^2 + b^2 &= 1, \quad a,b \in R.
  \end{aligned}
\end{equation}

This can also be written as a unitary transformation matrix:

\begin{equation}
  \begin{aligned}
    H''_{\text{EY}} &= \adj{R} H'_{\text{EY}} R, \\
    R &= 
      \begin{bmatrix}
        a  & 0  & b & 0 \\
        0  & a  & 0 & b \\
        -b & 0  & a & 0 \\
        0  & -b & 0 & a \\
      \end{bmatrix}.
  \end{aligned}
\end{equation}

The two subsequent transformations can be reduced to a single unitary transformation.

\begin{equation}
  \begin{aligned}
      H_{\text{EY,transformed}} &= \adj{U} H_{\text{EY}} U, \\
      U &= P R =
        \begin{bmatrix}
          a  & 0  & b & 0 \\
          0  & -b & 0 & a \\
          -b & 0  & a & 0 \\
          0  & a  & 0 & b \\
        \end{bmatrix}.
  \end{aligned}
\end{equation}

With appropriate $a$ and $b$ values, the off-diagonal blocks can be completely cancelled out.

\begin{equation}
  \begin{aligned}
    \Delta'_k &= \sqrt{\Delta_k^2 + 4 L_{2,k}^2}, \\
    a_k       &=                \sqrt{\frac{\Delta'_k+\Delta_k}{2\Delta'_k}}, \\
    b_k       &= \sign(L_{2,k}) \sqrt{\frac{\Delta'_k-\Delta_k}{2\Delta'_k}}.
  \end{aligned}
\end{equation}

After applying the transformation we get the following Hamiltonian:

\begin{widetext}
  \begin{equation}
    H_{\text{EY,transformed}} =
      \kbordermatrix{
                     & A'\uparrow              & A'\downarrow            & B'\uparrow              & B'\downarrow \\
        A'\uparrow   & 1/2(\Delta_k-\Delta'_k) & L_{1,k}                 & 0                       & 0 \\
        A'\downarrow & \conj{L}_{1,k}          & 1/2(\Delta_k+\Delta'_k) & 0                       & 0 \\
        B'\uparrow   & 0                       & 0                       & 1/2(\Delta_k-\Delta'_k) & L_{1,k} \\
        B'\downarrow & 0                       & 0                       & \conj{L}_{1,k}          & 1/2(\Delta_k+\Delta'_k)
      }.
    \label{eq:H_EY_transformed}
  \end{equation}
\end{widetext}

The resulting Hamiltonian can be treated similarly to Eq.~\ref{eq:H_EY_renamed},
  the only difference is that the instances of $\Delta_k$ have to be replaced with $\Delta'_k$ in the effective magnetic fields.

\section{An alternative derivation of the EY and GEY results using a DP consideration}

We derived in the main text that the EY model can be considered as being under the action of built-in SOC related magnetic fields and a much larger, band-gap related Zeeman field. This allows us to provide the schematics of spin relaxation for the EY and GEY regimes with a simple consideration.

\begin{figure}[htp]
\begin{center}
\includegraphics[width=.7\columnwidth]{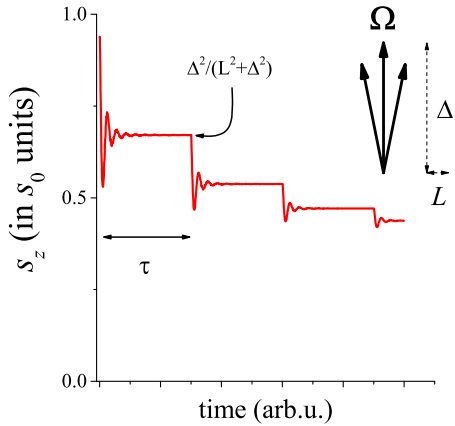}
\caption{Schematics of the time evolution of the $s_z$ component of the pseudospin-ensemble for the usual EY scenario when described in the DP framework. Note a rapid dephasing to an equilibrium value until the next momentum scattering occurs for the electron ensemble. Inset shows schematically the Larmor vector distribution for this case. The spin polarization tends to zero for long times.}
\label{EY_description}
\end{center}
\end{figure}

The situation for the conventional EY is depicted in Fig. \ref{EY_description}. We consider a pseudospin ensemble with different $\bm{k}$ vectors at $t=0$ and pseudospins polarized along the $z$ axis, i.e. $s_z(t=0)=s_0$. The Larmor precession vectors are such (according to Eq. \eqref{eq:Larmor}) that these describe a rotation nominally around the $z$ axis (this does not contribute to pseudospin-decay) with some spread in the vector directions in the $x-y$ plane. The latter components are much smaller as $L_{k}\ll \Delta_{k}$. The spread in the Larmor precession vector directions will cause a dephasing of the pseudospin polarization. However, a sizeable $z$ component remains and a geometric consideration shows it to be $s_{\text{dephased}}=s_0\frac{\Delta^2}{L^2+\Delta^2}$. The pseudospin ensemble retains this polarization until a consecutive momentum scattering randomizes the $\bm{k}$-vectors and the dephasing restarts. This means that per momentum scattering, the pseudospin polarization decay is $\frac{L^2}{L^2+\Delta^2}\approx \frac{L^2}{\Delta^2}$. As mentioned above, the $z$ component of the pseudospin decays together with the usual spin. 

It is interesting to note that the initial drop of the ensemble magnetization to $\frac{\Delta^2}{L^2+\Delta^2}$ is the same that the spin system experiences in the quantum mechanical description when the SOC is switched on from zero to a finite value: then, the up/down spin admixtures lower the spin value according to the wave functions in Eq. \eqref{Eq:EYmixed_states}.

To obtain the spin-relaxation time, we consider the time dependence of the spin ensemble magnetization in the following. After a time $\tau$ the value of $s_z$ reads:

\begin{gather}
s_z(t+\tau)\approx\frac{\Delta^2}{L^2+\Delta^2}\cdot s_z(t).
\end{gather}

The reduction of $s_z$ equals $\Delta s_z=-\frac{L^2}{L^2+\Delta^2}\cdot s_z(t)$. The finite difference, $\Delta s_z\approx \frac{\text{d}s_z}{\text{d}t}\tau$, yields the differential equation for $s_z(t)$ as:

\begin{gather}
\frac{\text{d}s_z}{\text{d}t}=-\frac{L^2}{\left(L^2+\Delta^2\right)\tau}s_z.
\end{gather}

Its solution yields $s_z(t)=s_z(0)e^{-t/\tau_{\text{s}}}$ with 

\begin{gather}
\frac{1}{\tau_{\text{s}}}=\frac{L^2}{L^2+\Delta^2}\frac{1}{\tau}.
\label{eq:approximated_Elliott} 
\end{gather}

In the limit of of low SOC, $L\ll \Delta$, it returns the usual Elliott formula. However, it is interesting to note that Eq. \eqref{eq:approximated_Elliott} also contains the solution when $L$ is significant, a result which was derived in Ref. \onlinecite{KissSciRep}.

To obtain the result in the GEY regime, we consider the dynamics of an individual pseudospin according to:

\begin{gather}
\frac{\text{d}\bm{s}}{\text{d}t}=-\bm{s} \times \bm\Omega_k,
\label{eq:spindynamics}
\end{gather}
where the minus sign is due to the negative electron charge and $\bm\Omega_k$'s are given in Eq. \eqref{eq:Larmor}. The pseudospins precess by only a small angle between two consecutive momentum scattering due to the short $\tau$ since $(L,\,\Delta)\ll \Gamma$. Thus Eq. \eqref{eq:spindynamics} can be linearized and the change in the $s_z$ component during a short time $\tau$ for a state $A$ reads (the result is obtained for a $B$ state in a similar manner):

\begin{gather}
\Delta \bm{s} =-\frac{2\tau}{\hbar}
\begin{bmatrix}
          -s_y\Delta/2-s_z\Im[L_k]\\
          -s_z \Re[L_k]-s_x\Delta/2\\
					s_x \Im[L_k]+s_y \Re[L_k].
\label{eq:spindynamics_linearized}
        \end{bmatrix}
\end{gather}

Eq. \eqref{eq:spindynamics_linearized} means that the change in the $z$ component is unaffected by the $\Delta$ term, i.e. the time evolution of the $z$ component behaves as if only the SOC related fields were present.

Then, the DP theory can be applied with the result in Eq. \eqref{Eq:DPresult} with the $\mathcal{L}$ substituted by $L$ which leads to the generalized EY result in the $\Delta\ll \Gamma$ limit as:

\begin{gather}
\Gamma_{\text{s}}=\frac{L^2}{\Gamma}
\end{gather}

\section{The Monte Carlo results of spin-relaxation rate for the EY modell with an isotropic SOC}

The main text described that spin-relaxation time in MgB$_2$ can be quantitatively reproduced using our approach. The assumptions of the calculation are the presence of a single (i.e. $\bm{k}$-independent) band-band separation value ($\Delta$) and an isotropic 4-band EY model (2 kinetic bands). Isotropy means that its Hamiltonian is invariant to spatial rotations. These are quite significant simplifying assumptions, however, the agreement between the calculation and the experimental data proves that this model is indeed adequate.

In this isotropic case, it is sufficient to specify the Hamiltonian for a single $\vec{k}_0$ point on the spherical Fermi-surface. The matrix elements are obtained for an arbitrary point by transforming this Hamiltonian by means of rotations. For convenience we specify this given $\vec{k}_0$ wave vector as the ``north pole'' on the Fermi-sphere ($\vec{k}_0 = (0,0,k_\mathrm{F})$). We are interested in the matrix elements of the Hamiltonian such as:

\begin{equation}
  H_{\alpha,\sigma;\alpha',\sigma'}(\vec{k})=
    \bra{ \vec{k}, \alpha, \sigma }
    H
    \cat{ \vec{k}, \alpha', \sigma'},
\end{equation}
where $\alpha$ and $\sigma$ are the band and spin indices, respectively.

The matrix elements between wave functions with different $\vec{k}$ and $\vec{k}'$ are $0$.
Treating the $\alpha, \sigma$ pair as a single index, the Hamiltonian can be conveniently treated as a $\vec{k}$ dependent $4\times4$ matrix.
For the case of a retained inversion symmetry, the most general form of this matrix reads:

\begin{equation}
  \mat{H} =
    \kbordermatrix{
                    & 1\uparrow      & 1\downarrow & 2\uparrow      & 2\downarrow \\
        1\uparrow   & 0              & 0           & L_{2,k}        & L_{1,k} \\
        1\downarrow & 0              & 0           & \conj{L}_{1,k} & -L_{2,k} \\
        2\uparrow   & L_{2,k}        & L_{1,k}     & \Delta_{k}     & 0 \\
        2\downarrow & \conj{L}_{1,k} & -L_{2,k}    & 0              & \Delta_{k}
    }.
    \label{eq:H_EY_gen}
\end{equation}

Rotations around the $z$ axis leave the $\vec{k}_0$ vector invariant.
The generator of rotations around the $z$ is $J_z=L_z+S_z$.

\begin{equation}
  \begin{aligned}
    \left[H, J_z\right] &= 0, \\
    \bra{ \vec{k}_0, \alpha, \sigma }
    \left[H, J_z\right]
    \cat{ \vec{k}_0, \alpha', \sigma' } &= 0.
  \end{aligned}
\end{equation}

Since we have two non-degenerate kinetic bands, we can assume that the angular momentum is quenched. It yields:

\begin{gather}
    \bra{ \vec{k}_0, \alpha, \sigma }
    \left[H, S_z\right]
    \cat{ \vec{k}_0, \alpha', \sigma' } = 0, \nonumber \\
    \left[H, S_z\right]_{\alpha,\sigma;\alpha',\sigma'}(\vec{k}_0) =
      \begin{bmatrix}
        0 & 0 & 0 & -L_{1,\vec{k}_0} \\
        0 & 0 & \conj{L}_{1,\vec{k}_0} & 0 \\
        0 & -L_{1,\vec{k}_0} & 0 & 0 \\
        \conj{L}_{1,\vec{k}_0} & 0 & 0 & 0 \\
      \end{bmatrix}, \nonumber \\
    L_{1,\vec{k}_0} = 0.
\end{gather}

This means that for the $k_0$ point, the SOC matrix elements are described by a single real parameter: $L=L_{2,\vec{k}_0}$.

A general rotation transformation is described by the following operation:

\begin{equation}
  R = e^{-i \vec{J} \vec{a} \varphi /\hbar} = e^{-i \vec{L} \vec{a} \varphi /\hbar } e^{-i \vec{S} \vec{a} \varphi/\hbar},
\end{equation}

\noindent where $\vec{a}$ and $\varphi$ denote the axis and angle of the rotation, respectively. Clearly, the rotation operation acts separately on the kinetic and spin parts of the wave function, i.e.:

\begin{equation}
  R \cat{ \vec{k}, \alpha, \sigma } 
    = (e^{-i \vec{L} \vec{a} \varphi/\hbar} \cat{ \vec{k}, \alpha })
      \otimes (e^{-i \vec{S} \vec{a} \varphi/\hbar } \cat{ \sigma }).
\end{equation}

For any $\vec{k}\ne \vec{k}_0$, we can define a single operation that rotates $\vec{k}_0$ to $\vec{k}$.

\begin{equation}
  \begin{aligned}
    \vec{a}_{\vec{k}_0,\vec{k}} &=
      \frac{
        \vec{k}_0 \times \vec{k}
      }{
        \left| \vec{k}_0 \times \vec{k} \right|
      }, \\
    \varphi_{\vec{k}_0,\vec{k}} &= \sin^{-1}( \left| \vec{k}_0 \times \vec{k} \right| ), \\
    R_{\vec{k}_0,\vec{k}} &= e^{-i \vec{J} \vec{a}_{\vec{k}_0,\vec{k}} \varphi_{\vec{k}_0,\vec{k}}/\hbar}.
  \end{aligned}
\end{equation}

The spatial rotations are symmetries of the kinetic part of the Hamiltonian thus the rotation does not mix wave functions from different bands, i.e.:

\begin{equation}
  \cat{ \vec{k}, \alpha } = R_{\vec{k}_0,\vec{k}} \cat{ \vec{k}_0, \alpha}.
\end{equation}

Our Hamiltonian is symmetric to these rotations, therefore:

\begin{equation}
  \begin{aligned}
    \left[H,R_{\vec{k}_0,\vec{k}}\right] &= 0, \\
    H &= R_{\vec{k}_0,\vec{k}}H R_{\vec{k}_0,\vec{k}}^{-1} .
  \end{aligned}
\end{equation}

The only non-zero matrix elements in the $\vec{k}$ vector read:

\begin{widetext}
\begin{equation}
  \begin{aligned}
    \bra{ \vec{k}, \alpha, \sigma }
      H
    \cat{ \vec{k}, \alpha', \sigma' } &=
    \bra{ \vec{k}, \alpha, \sigma }
      R_{\vec{k}_0,\vec{k}}H R_{\vec{k}_0,\vec{k}}^{-1}
    \cat{ \vec{k}, \alpha', \sigma' } \\
    \bra{ \vec{k}, \alpha, \sigma }
      H
    \cat{ \vec{k}, \alpha', \sigma' } &=
    \bra{ \vec{k}_0, \alpha, \sigma }
      e^{-i \vec{S} \vec{a}_{\vec{k}_0,\vec{k}} \varphi_{\vec{k}_0,\vec{k}}/\hbar}
      H
      e^{i \vec{S} \vec{a}_{\vec{k}_0,\vec{k}} \varphi_{\vec{k}_0,\vec{k}}/\hbar}
    \cat{ \vec{k}_0, \alpha', \sigma' }, \\
    H_{\alpha,\sigma;\alpha',\sigma'}(\vec{k}) &=
      \sum_{\sigma_1,\sigma_2}
        \left[
          e^{-i \vec{S} \vec{a} \varphi/\hbar}
        \right]_{\sigma,\sigma_1}
        H_{\alpha,\sigma_1;\alpha',\sigma_2}(\vec{k}_0)
        \left[
          e^{i \vec{S} \vec{a} \varphi/\hbar}
        \right]_{\sigma_2,\sigma'}.
  \end{aligned}
\end{equation}
\end{widetext}

The spin rotation can be calculated using matrix exponential of $2\times2$ matrices.
This allows us to arrive at the final SOC values at the $\vec{k}=(k_x,k_y,k_z)$ point:

\begin{equation}
  \begin{aligned}
    L_{1,\vec{k}} &= \frac{-L k_x - i L k_y}{k_\mathrm{F}}, \\
    L_{2,\vec{k}} &= \frac{ L k_z }{k_\mathrm{F}}.
  \end{aligned}
\end{equation}

This remarkably simple and symmetric result allows the calculation of the SOC Hamiltonian for any $\vec{k}$ points. This, together with the above transformation of the EY Hamiltonian to the DP problem and the corresponding Monte Carlo method described in the main text, allows the calculation of spin-relaxation times for this isotropic system.

\begin{figure}%[htp]
\begin{center}
\includegraphics[width=.8\columnwidth]{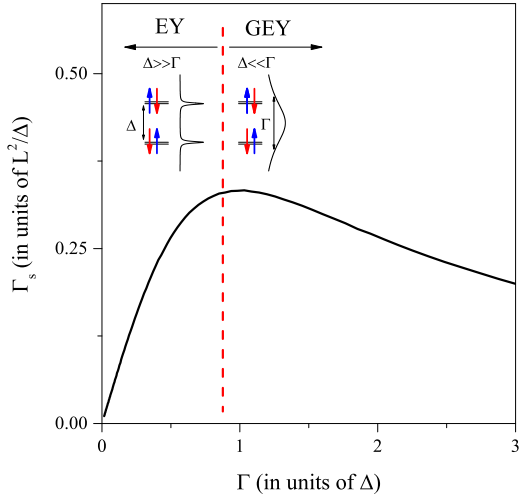}
\caption{Spin-relaxation rate, $\Gamma_{\text{s}}$ calculated for an isotropic SOC with varying momentum relaxation rate, $\Gamma$. The schematics of the band structure is also shown and the EY and GEY regimes are indicated. Note that the crossover between the two regimes is rather smooth.}
\label{fig:FigSM1_MgB2_calc}
\end{center}
\end{figure}

Fig. \ref{fig:FigSM1_MgB2_calc}. shows the universal curve which is obtained for the spin-relaxation of a metal with inversion symmetry and isotropic SOC. The details of the calculation are outlined in the main text and essentially mimic the method in Ref. \onlinecite{SzolnokiSciRep}. The curve is well fitted by 

\begin{equation}
\Gamma_{\text{s}}\left(\Gamma\right)=\frac{2}{3}\frac{\Gamma}{1+\Gamma^2},
\end{equation}
where $\Gamma_{\text{s}}$ is measured in units of $\text{units of L}^2/\Delta]$ and $\Gamma$ is measured in units of $\Delta$. The 2/3 prefactor depends on the way the SOC matrix elements are defined and the original Elliott result allows for the presence of similar factors (usually denoted by $\alpha$).

%\bibliographystyle{apsrev}

%\bibliography{Tubes2011June_new}

\end{document}